\documentclass[superscriptadress,twoside,onecolumn]{revtex4}

\usepackage{eurosym}
\usepackage{bm}
\usepackage[utf8]{inputenc}
\usepackage{amssymb,theorem,xspace}
\usepackage{float}
\usepackage{amsmath} 
\usepackage{mathrsfs}
\usepackage{graphicx}
\usepackage{xcolor}
\usepackage{caption}
\usepackage{subcaption}
\usepackage{booktabs}
\usepackage{cleveref}
\begin{document}
\title{Symplectic Pauli-Schrödinger Equation for the Quark-Antiquark Interaction System}
\author{R.R. Luz}
\email{renatoluzrgs@gmail.com}
\affiliation{International Center for Physics, Instituto de Física, Universidade de Brasília, Brasília, DF, Brazil }

\affiliation{Departamento de Física, Universidade Estadual do Maranhão, São Luís, MA, Brazil}

\author{R.A.S. Paiva}
\email{rendisley@gmail.com}
\affiliation{FCTE/Universidade de Brasília, Brasília, DF, Brazil }

\author{G.X.A. Petronilo}
\email{gustavopetronilo@gmail.com}
\affiliation{
Universidade Federal do Pará, Salinópolis,PA, Brazil }

\author{A.E. Santana}
\email{a.berti.santana@gmail.com}
\affiliation{International Center for Physics, Instituto de Física, Universidade de Brasília, Brasília, DF, Brazil }

\author{T.M. Rocha-Filho}
\email{dimarti@unb.br}
\affiliation{International Center for Physics, Instituto de Física, Universidade de Brasília, Brasília, DF, Brazil }

\author{R.G.G. Amorim}
\email{ronniamorim@gmail.com}
\affiliation{International Center for Physics, Instituto de Física, Universidade de Brasília, Brasília, DF, Brazil }
\affiliation{Canadian Quantum Research Center, Canada}

\begin{abstract}
 We investigate the quantum behavior of a quark-antiquark  bound system under the influence of a magnetic field within the symplectic formulation of quantum mechanics. Employing a perturbative approach, we obtain the ground and first excited states of the system described by the Cornell potential, which incorporates both confining and non-confining interactions. After performing a Levi-Civita mapping in phase space, we solve the time-independent symplectic Pauli–Schrödinger-type equation and determine the corresponding Wigner function. Special attention is given to the observation of the confinement of the quark-antiquark, that is revealed in the phase space structure. Due to the presence of spin in the Hamiltonian, the results reveal that the magnetic field enhances the non-classicality of the Wigner function, signaling stronger quantum interference and a departure from classical behavior.  The experimental mass spectra  is used to estimate the intensity of the external field, leading to a value that is in order of the transient magnetic field measured in non-central heavy-ion collisions at RHIC and LHC.
\end{abstract}

\maketitle

\paragraph*{Authors' Note:} The present work can be found in the arxiv repository with the following identification number \url{	arXiv:2507.20045 [quant-ph]}.

\section{Introduction}
\label{sec:Introduction}

Particle physics aims to understand the fundamental laws of physics from basic structures, described by
the standard model that addresses for instance fundamental interactions: the strong, weak and electromagnetic
forces~\cite{Godfrey,bali2001qcd,Grosse,fulcher1993,lucha1989phen,Gupta}. In more than half a century, particle accelerators
allowed to  access experimentally higher energies leading to the discovery of fundamental particles such as quarks~\cite{thomson,Marciano,mann,abu, abusha,AbuSh2023,gasiorowicz,griffiths,aitchison}.
These are particles confined in nature, with the gluons being the interacting field. This is the mechanism giving rise to hadrons~\cite{Omugbe,Brambillaa}.

Different approaches have been developed to obtain predictions from the particle standard model, such as lattice calculations in quantum chromodynamics (QCD),
effective field theory treatments, QCD sum rules and potential models~\cite{RL_1}.
Among the latter, a well known  quark interaction  is described by the Cornell (or funnel) potential. This model successfully takes into account the two crucial features of QCD, namely, the asymptotic freedom
and the quark confinement~\cite{Kumar,Mansour}. This is  an important  tool for
examining the transition between the confined and deconfined phases of matter~\cite{Bruni,belich,shady,Koothottil}. Beyond particle physics, the Cornell  potential is of interest in other areas of physics such as nuclear physics, condensed matter
and cosmology~\cite{Tajik,Cottingham,AS_3,Hassanabadi,Lombard,Soloviev,Chand,Montigny,Spallucci}.

For a proper choice of units and covering a sector of QCD, the phenomenological Cornell potential is given by~\cite{Eich,Eichten,MAIRECHE,Bukor}
\begin{equation}
   V(r)=-\frac{\alpha}{r}+\beta r, 
	\label{cornpot}
\end{equation}
where $r$ is the inter-quark distance, $\alpha >0$ is the quark-gluon coupling constant and $\beta$ stands for the confinement
constant~\cite{Mab,Ciftci,Hamzavi,Chung}. In other words, the first ( a Coulomb-like) term on the right side of Eq.~(\ref{cornpot}) describes
the one-gluon exchange between the quark-antiquark interaction; a dominant effect at shorts distances. The second (linear) term accounts
for the quark confinement at large inter-quark separation~\cite{Mutuk2019,Ferkous2013,Sameer2013,rluz2023frac}. In high energy, such a potential has been used for studying heavy quarkonia. This includes the calculation of masses of heavy quarkonium states
by considering a quark-gluon plasma, as consisting of bound states of quarks and gluons~\cite{Koothottil,As_4,Aydin,Ramazanov,Rannu}.

From a theoretical point of view, quarkonium systems such  as $b\overline{b}$,  $c\overline{c}$ and $\overline{b}c$ have been analyzed by using both relativistic and non-relativistic
wave equations, which are  solved by employing different approaches~\cite{vega}. This is the case, for instance, of considering the following aspects:  
an extra dimension, as an extension of the Cornell potential~\cite{Lahkar};  an inclusion of a harmonic oscillator term for the Dirac equation~\cite{ABUSHADY};  
  a Lie algebraic method of quasi-exactly solvable systems~\cite{Panahi};
the asymptotic iteration method~\cite{Kumar,Abdel-Karim,Hall}; the mass spectrum of bottomonium has been also analyzed by using Gaussian wave functions~\cite{Kher}; bound states of the finite temperature-dependent Schrödinger equation for an extended Cornell potential were also calculated~\cite{Ahmadov}, such that the mass spectra of heavy quarkonia and the $B_{c}$ meson at both zero and nonzero temperatures were derived; and the method supersymmetric expansion algorithm-(SEA) has been utilized to obtain  analytical solutions, leading to relativistic corrections. Hence, it was possible to extract general qualitative and quantitative predictions for the bottomonium and charmonium spectrum which nicely agree with  experimental data~\cite{Rodriguez}.

The Schrödinger equation with the Cornell potential in an Euclidian non-commutative space has been analyzed~\cite{AJamel}. In this framework, it was obtained analytical solutions and spectral properties  for the deformed Schrödinger, Dirac and Klein-Gordon equations. These developments utilize the Bopp shift method and perturbation theory in the phase-space Wigner function~\cite{AbMaireche,Imane,AbdelMaire}.
Along the same direction, 
solutions of the symplectic Schrödinger equation provided, as an instance, characteristics of the confinement-deconfinement due to the structure of the quantum phase-space representation~\cite{RL_1}. More recently a generalized Cornell potential for
heavy quark interaction was considered in the context of the Nikiforov-Uvarov method, with fractional derivatives in the
framework of the symplectic quantum mechanics~\cite{renato2024frac}.

As an Euclidian potential, quite often the Schrödinger equation for a boson is used to analyze the spectrum and the evolution of the quark system. Nevertheless, since the quark is a fermion the (non-relativistic) Pauli-Schrödinger equation should be of interest. Indeed, this aspect was explored to analyze  the Casimir effect for the confined quark-antiquark system, near the region of $V(r)=0$~\cite{petronilo}, in Eq.~(\ref{cornpot}). In this case, the symplectic representation of quantum mechanics was used, providing improvements in the results on confinement due to the presence of spin~\cite{AdemirESant}.  

The symplectic representation of Schrödinger equation is intimately associated with the Wigner function formalism, which has been widely used in the literature, as in  particle systems~\cite{Jana}. The Wigner function can be reconstructed from
optical homodyne tomography~\cite{Beck,Schiller} or directly sampled point-by-point by using photon counting and displacement
techniques~\cite{Ferry,Banaszek,Huang}. It is a powerful tool for determining the statistical characteristics of quantum states
such as non-classicality and aspects of chaoticity. These properties are crucial  not only in the fields of quantum computing and
quantum information but also play a significant role in particle physics~\cite{renato2024frac,Jana,Lorce,Ojha,Weinbub}.

In the present work, we study the quantum-mechanical problem of a  quarkonium  moving in the Cornell potential, by using the symplectic quantum mechanics approach including the spin effect. Our main goals and  new contributions for the problem are included in the following aspects. (a)  We use symmetry to derive the symplectic representation of the Pauli-Schrödinger equation.  (b) Aspects of the confinement are investigated considering experimental results. (c) The experimental spectrum is  considered in order to estimate the intensity of the external field. (d) The emergency of non-classicality in the quarkonium states are analyzed as a function of the external field.  In order to solve the symplectic Pauli-Schrödinger equation, the Levi-Civita map  and the standard perturbation theory up to first and second order are used. This leads to the Wigner function for the quark-antiquark bound state due to a Cornell potential and a magnetic field.

It is important to mention that in the Hamiltonian, we have included the interaction of the spin with the magnetic field, $B$, only. In this way we have been able to estimate the value of $B$ in a satisfactory agreement with experiments. The spin-orbit term has been studied in phase space but in a different context and for another physical problem~\cite{wozniak2025phase}. In that case a symplectic structure of the Wigner function is also associated with the spin by using a $S^2$ topology.  This procedure is interesting for arbitrary interacting system since  it is a hard procedure to  introduced spin and  gauge fields with the Wigner function or equivalently for the density matrix. 
In the present work, a symplectic Hilbert space is explored as the representation space for the Galilei Lie group. As a consequence, the spin 1/2 representation arises naturally and the coupling with external field is carried out by a covariant derivative. It is worth mention different formulations of the Schrödinger equation in phase space~\cite{torres1993quantum,chruscinski2005wigner}. In the present work we use a representation which is directly associated with the Wigner function~\cite{AdemirESant}.

The work is organized as follows. In Section~\ref{sqm}, in order to fix the notation,  we discuss basic aspects of the  symplectic Schrödinger equation,
and the relation between the phase-space wave function (a quasi-amplitude of probability) and Wigner function (a quasi-distribution of probability). Section~\ref{FSE} presents the quark-antiquark
description using the Symplectic Pauli-Schrödinger equation,
with the Cornell potential in the presence of an external magnetic field.
Section~\ref{dr} brings the discussions for the  results.
Some final concluding remarks are presented in Section~\ref{fr}.

\section{Symplectic Quantum Mechanics: brief review  and notation}
\label{sqm}
In this section, we present some few aspects of  the non-relativistic symplectic formulation of quantum mechanics for boson are presented~\cite{AdemirESant} and fix the notation.
Consider the set $\Gamma$ of the points with coordinates $(q,p)$, $q,p\in\mathbb E$, where $\mathbb E$ is the  Euclidean space.
When equipped with the symplectic 2-form $w=dq\wedge dp$, $\Gamma$ is denominated  phase space. 

A Hilbert space in
$\Gamma$, say $\mathcal{H}(\Gamma)$, is formed by the set of $C^\infty$ functions $\phi(q,p)$ with the property
$
\int  dqdp\,\phi ^{\dagger}(q,p)\phi (q,p)< \infty.
$
A basis in $\mathcal{H}(\Gamma)$ is given by 
$|q,p\rangle$, with dual $\langle q,p|$, such that   
$
\phi(q,p)= \langle q,p | \phi\rangle,
$
and
$
\phi^{\dagger}(q,p)=\langle\phi|q,p\rangle.
$
Unitary mappings $U(\alpha )$ in $\mathcal{H}(\Gamma )$ are defined by $U(\alpha )=\exp(-i\alpha \widehat{A})$, where the generator of symmetry $\widehat{A}$ is given by 
\begin{equation}
\widehat{A} = A(q,p)\star =A(q,p)\exp \left[ \frac{i\hbar }{2}
\left(\frac{\overleftarrow{\partial }}{\partial q}
\frac{\overrightarrow{\partial }}{\partial p}-\frac{\overleftarrow{\partial }}{\partial p}
\frac{\overrightarrow{\partial }}{\partial q}\right)\right]=
 A\left(\widehat{Q},\widehat{P}\right), \label{g000}
\end{equation}
where we have defined 
\begin{eqnarray}
\widehat{P} &=&p\star =p-\frac{i\hbar}{2}\partial_{q},
\label{g001}\\
\widehat{Q} &=&q\star =q+\frac{i\hbar}{2}\partial_{p}.
\label{g1}
\end{eqnarray}
Here, the following notation for derivatives is used, i.e., $\partial_{q}\equiv\partial/\partial q$, $\partial_{p}\equiv\partial/\partial p$. The 
arrows in Eq.~(\ref{g000})  mean that the respective operator acts on a function at its left (right) 
according to the orientation of  the arrow $\leftarrow$ ($\rightarrow$). The star (or Weyl) product, $\star$, is defined by 
$\star\equiv\exp \left[ \frac{i\hbar }{2}\left(\frac{\overleftarrow{\partial }}{\partial q}
\frac{\overrightarrow{\partial }}{\partial p}-\frac{\overleftarrow{\partial }}{\partial p}
\frac{\overrightarrow{\partial }}{\partial q}\right)\right].$

From now on we will set units such that $\hbar=1$, the speed of light is $c=1$ and the electron charge $e=1$, unless necessary for clarifying the units.

In the symplectic framework for quantum mechanics the Heisenberg commutation relation is written as
$\left[ \widehat{Q}_j,\widehat{P}_k\right] =i\delta_{jk}$, with $j,k=1,2,3$. Then the operators $\widehat{Q}=(\widehat{Q}_1,\widehat{Q}_2,\widehat{Q}_3)$ and $\widehat{P}=(\widehat{P}_1,\widehat{P}_2,\widehat{P}_3)$ stand for the observables position and momentum, respectively. Considering the Galilei symmetries, the following operators are defined:
$
\widehat{K}_{i} \equiv m\widehat{Q}_{i}-t\widehat{P}_{i},\hspace{3mm}
\widehat{L}_{i} \equiv \epsilon _{ijk}\widehat{Q}_{j}\widehat{P}_{k},\hspace{3mm}
\widehat{H} \equiv \widehat{P}^{2}/2m. \,\,\, i=1,2,3.
$
Here $m$  stands for the mass of a particle. The operators  $\widehat{P},\widehat{K},\widehat{L}$ and $\widehat{H}$
are then the generators of Galilei symmetries, standing  for spatial translation (linear momentum), Galilean boosts, rotation (the angular momentum), and time translation (the free Hamiltonian), respectively.

The time-translation generator $\widehat{H}$  provides the time evolution of a wave function in phase space $\psi(q,p,t_0)$ from a initial time $t_0$ to a time $t$ by writing 
$
\psi(q,p,t)=e^{-i\widehat{H}(t-t_0)}\psi(q,p,t_0),
$
which is the formal solution of the  Schrödinger equation in phase space, i.e., 
\begin{equation}
i\partial_t \psi(q,p;t)=\widehat{H}(q,p) \psi(q,p;t),
\label{SSEqI}
\end{equation}
also called the symplectic Schrödinger equation~\cite{oliveira2004}.
The physical interpretation of this formalism is given by the relation of the phase-space wave function $\psi(q,p,t)$
with the Wigner function, $f_W(q,p,t)$~\cite{oliveira2004,paiva2018,dessano,paiva2020,Martins}, that is 
$$
f_W(q,p,t)=\psi(q,p,t)\star\psi^\dagger(q,p,t).
$$
Since $f_W(q,p,t)$ is a quasi-distribution of probability,  $\psi(q,p,t)$ is interpreted as a quasi-amplitude of probability~\cite{RL_1,paiva2020,campos2017,campos2018}. For our proposals here, it is important to write the eigenvalue version of Eq.~(\ref{SSEqI}), the steady state equation, i.e., 
\begin{equation}
(H(q,p) - E)\star \psi(q,p)=0,
\label{SSEqII}
\end{equation}
where $E$ is the eigenvalue of $\widehat{H}$.

All the results reviewed above are directly
generalized for $N$-dimensions. It is important also to note that Eq.~(\ref{SSEqI}) describes particle with spin zero. In the next section  this approach is generalized to consider particles with spin 1/2 in the presence of an external field.

\section{Quark-antiquark bound states in symplectic manifold}
\label{FSE}
This section is presented in three parts. The first, Subsection A, the construction of the general formalism is addressed. That is, the bound states of the quark-antiquark system in  the symplectic formulation are derived, by considering the spin effect. The perturbative results are analyzed in Subsections B and C. 

\subsection{General formalism}
Initially, let us consider a non-relativistic bound state of a massive quark-antiquark system in an external magnetic field. At this level, the quark-antiquark system is described in terms of its effective
internal dynamics in the meson rest frame. For the equal-flavor quarkonia considered here, the quark and antiquark have equal constituent masses, $m_q=m_{\bar q}$, and opposite electromagnetic couplings, $+e$ and $-e$, respectively. The corresponding non-relativistic two-constituent Pauli Hamiltonian,
obtained through minimal electromagnetic coupling~\cite{Sakurai2020,Alford2013CharmoniaAB}, may be written as
\begin{equation}
\begin{aligned}
H={}
\frac{
\left[\mathbf p_q-e\mathbf A(\mathbf r_q)\right]^2
}{2m_q}
+
\frac{
\left[\mathbf p_{\bar q}+e\mathbf A(\mathbf r_{\bar q})\right]^2
}{2m_q}-
\frac{ge}{4m_q}
\left(
\boldsymbol{\sigma}_q-\boldsymbol{\sigma}_{\bar q}
\right)\cdot\mathbf B
+V(r),
\end{aligned}
\label{hp}
\end{equation}
where $r=|\mathbf r_q-\mathbf r_{\bar q}|$ is the inter-quark distance and $V(r)$ is the interaction potential. Higher-order relativistic and additional spin-dependent corrections are not retained in the effective description considered here.

For the internal motion in the meson rest frame, we use
$\mathbf r_q=\mathbf r/2$, $\mathbf r_{\bar q}=-\mathbf r/2$,
$\mathbf p_q=\mathbf P$, and
$\mathbf p_{\bar q}=-\mathbf P$. The corresponding reduced mass is
$m=m_q/2$. For a spatially uniform magnetic field in the $z$
direction, $\mathbf B=B\hat{z}$, the symmetric gauge is written as
$\mathbf A(\mathbf r)=\mathbf B\times\mathbf r/2$. Since the vector
potential is linear in the coordinates, one has
$\mathbf A(\mathbf r_q)=\mathbf A(\mathbf r)/2$ and
$\mathbf A(\mathbf r_{\bar q})=-\mathbf A(\mathbf r)/2$.

The kinetic contributions of the two constituents are therefore given by
\begin{equation}
\begin{aligned}
&\frac{1}{2m_q}
\left[
\left(\mathbf P-\frac{e}{2}\mathbf A\right)^2
+
\left(\mathbf P+\frac{e}{2}\mathbf A\right)^2
\right]
\\
&=
\frac{1}{2m_q}
\left[
2\mathbf P^2
+\frac{e^2}{2}\mathbf A^2
-\frac{e}{2}
\left(\mathbf P\!\cdot\!\mathbf A+\mathbf A\!\cdot\!\mathbf P\right)
+\frac{e}{2}
\left(\mathbf P\!\cdot\!\mathbf A+\mathbf A\!\cdot\!\mathbf P\right)
\right]
\\
&=
\frac{\mathbf P^2}{m_q}
+
\frac{e^2\mathbf A^2}{4m_q}.
\end{aligned}
\end{equation}
The terms linear in the vector potential cancel identically. Hence, the internal orbital magnetic coupling proportional to $\mathbf B\cdot\mathbf L$ does not survive for the equal-mass, oppositely charged quark-antiquark system considered here. This cancellation follows from the constituent structure of the effective
model and is not introduced as an approximation based on mathematical or computational convenience. On the other hand, the quadratic diamagnetic contribution remains.

The spin interaction in Eq.~\eqref{hp} depends on the difference between the quark and antiquark spin operators. Since the external magnetic field is directed along the $z$ axis, we define the effective spin projection as
$\sigma_z=(\sigma_{q,z}-\sigma_{\bar q,z})/2$, so that
$(\boldsymbol{\sigma}_q-\boldsymbol{\sigma}_{\bar q})\cdot\mathbf B=2\sigma_zB$.
For the spin channels considered in the present model, $\sigma_z=\pm1$. Taking $g\simeq2$, using $e=1$, and recalling that $m_q=2m$, the spin-field interaction becomes
$$
-\frac{ge}{4m_q}(2\sigma_zB)
=
-\frac{\sigma_zB}{2m}.
$$

Since the Levi-Civita transformation employed below is two-dimensional, the effective relative dynamics is considered in the $(x,y)$ plane, with $r=\sqrt{x^2+y^2}$. Using $\mathbf A^2=B^2(x^2+y^2)/4$, $m_q=2m$, and the natural units adopted in this work, the Hamiltonian becomes
\begin{equation}
H=
\frac{P_x^2+P_y^2}{2m}
+
\frac{B^2(x^2+y^2)}{32m}
-
\frac{\sigma_zB}{2m}
+
V(r).
\label{H2D}
\end{equation}

Using Eq.~(\ref{cornpot}) for the Cornell potential, the effective internal Hamiltonian is then written as
\begin{equation}
H=
\frac{P_x^2+P_y^2}{2m}
+
\frac{B^2(x^2+y^2)}{32m}
-
\frac{\sigma_zB}{2m}
-
\frac{\alpha}{r}
+
\beta r .
\label{H2D1}
\end{equation}

Equation~\eqref{H2D1} defines the effective internal Hamiltonian adopted in the following phase-space analysis. Center-of-mass dynamics, relativistic corrections, and additional spin-dependent interactions lie outside the scope of the present effective model. 

In order to address the Coulomb-like term in the Cornell potential, the Levi-Civita mapping (or Levi-Civita transformation) is considered. The transformation introduces the auxiliary canonical variables
$(q_1,q_2,p_1,p_2)$ associated with the two-dimensional relative dynamics. The Levi-Civita mapping is introduced here by~\cite{RL_1,paiva2020,campos2017,celletti2006} 
\begin{align}
x &= q_1^2-q_2^2,\label{bohlin2} \\
y &=2q_1q_2, \label{bohlin3}\\
P_x &=\frac{p_1q_1-p_2q_2}
{2(q_1^2+q_2^2)},\label{bohlin4}\\
P_y &=\frac{p_2q_1+p_1q_2}
{2(q_1^2+q_2^2)}. \label{bohlin5}
\end{align}
Substituting Eqs.~\eqref{bohlin2}-\eqref{bohlin5} in
Eq.~\eqref{H2D1}, and using
$r=\sqrt{x^2+y^2}=q_1^2+q_2^2$ and $\sigma_z=\pm1$, we obtain a central result for the present work,
\begin{equation}
H=
\left[
\frac{p_1^2+p_2^2}
{8m\left(q_1^2+q_2^2\right)}
+
\frac{B^2\left(q_1^2+q_2^2\right)^2}{32m}
-
\frac{\pm B}{2m}
\right]
-
\frac{\alpha}{q_1^2+q_2^2}
+
\beta\left(q_1^2+q_2^2\right).
\label{phham}
\end{equation}  
The interpretation of this Hamiltonian follows the standard application of the Levi-Civita regularization in Kepler-type systems: after fixing the energy hypersurface, the original singular Hamiltonian is mapped into an auxiliary regularized Hamiltonian in the transformed variables. Therefore, Eq.~\eqref{phham} is not
associated with a new physical particle; it is a mathematical representation of the same quark-antiquark system. This procedure is useful for removing the Coulomb-like singularity and for performing the subsequent symplectic quantization.

Considering the hypersurface in phase space given by $H=E$, where $E$ is a constant, it leads to  
\begin{equation}
\left(\frac{p_1^2+p_2^2}{2m}\right)-\frac{\pm 2B}{m}\left(q_1^2+q_2^2\right) 
-4E\left (q_1^2+q_2^2\right) 
+\frac{B^2\left(q_1^2+q_2^2\right)^3}{8m}
+4\beta\left(q_1^2+q_2^2\right)^2-4\alpha=0.
\label{pauliS}
\end{equation}
It is important to emphasize that the Levi-Civita mapping is a canonical transformation~\cite{campos2017,celletti2006}. In this sense, the physical content of the Hamiltonian in  Eq.~\eqref{pauliS} is the same as in its quantum version. In order to obtain the symplectic quantization, we use Eq.~(\ref{SSEqII}) in this context, which is equivalent to multiply the right side of Eq.~\eqref{pauliS}  by $\star\Psi (q,p)$, leading to
\begin{equation}
 \left[ \left(\frac{p_1^2+p_2^2}{2m}\right)-\frac{\pm 2B}{m}\left(q_1^2+q_2^2\right) 
-4E\left (q_1^2+q_2^2\right) 
+\frac{B^2\left(q_1^2+q_2^2\right)^3}{8m}
+4\beta\left(q_1^2+q_2^2\right)^2-4\alpha \right]\star \Psi=0. \label{SSPS22}
\end{equation}

Looking for a perturbative approach, Eq.~(\ref{SSPS22}) is rewritten as
\begin{equation}
\left [\widehat{H}_{0} + \widehat{H}_{1} \right ]\Psi(q_{1},p_{1},q_{2},p_{2}) = 4\alpha\Psi(q_{1},p_{1},q_{2},p_{2}),
\label{pauli2}
\end{equation}
where 
$\widehat{H}_{0}$ and $\widehat{H}_{1}$ are the unperturbed and perturbed Hamiltonians, respectively, given by
\begin{eqnarray}
\widehat{H}_{0} & = & \frac{1}{2m}\left( p_{1}^{2}+p_{2}^{2}\right)\star
	-\left ( \frac{\pm 2B }{m}+4E \right )\left (q_{1}^{2}+q_{2}^{2}\right)\star,
 \label{H0}   \\
\widehat{H}_{1} & = & \frac{B^2\left(q_1^2+q_2^2\right)^3 \star}{8m}
+4\beta\left(q_1^2+q_2^2\right)^2\star.
\label{H1}
\end{eqnarray}
The field  $\Psi(q,p)$ is the two-component Pauli-Schrödinger spinor, written as 
\begin{equation}
    \Psi =\binom{\psi_1}{\psi_2} \label{Paulis1}.
\end{equation}

In Eq.~(\ref{H0}), the frequency, $\omega$, of the two harmonic oscillators is given by 
\begin{equation}
\omega^{2}= -\left ( \frac{\pm 4B }{m^2}+\frac{8E}{m} \right ). \label{frequen}
\end{equation}

It is important to notice that Eq.~(\ref{pauli2}) looks like an eigenvalue equation, although $\alpha$ is a constant due to the Cornell potential. Despite this fact, in order to find solutions of Eq.~(\ref{pauli2}), it is possible to consider $\alpha$ as an non-fixed constant. At the end of the calculations the original value of $\alpha$ will be recovered. In the following, perturbation procedures will carried out, initially with the zero order.

\subsection{Zero order solution}
 
Using the fact that the Hamiltonian $\widehat{H}_{0}$ is that of a two-dimensional isotropic harmonic oscillator, 
we solve the zero order equation $\widehat{H}_{0}\Psi^{(0)}=4\alpha_{n_1,n_2}^{(0)}\Psi^{(0)}$, leading to
\begin{equation}\label{bn12}
	\Psi(q_1,p_1,q_2,p_2)=\phi_{n_1}(q_1,p_1)\phi_{n_2}(q_2,p_2),
\end{equation}
where $\phi_n(q,p)$ is the stationary state of the one-dimensional harmonic oscillator with quantum number $n$. 
Here, the quantum numbers $(n_1,n_2)$ label the occupation numbers of the two oscillator-like modes associated with the Levi-Civita variables $(q_1,q_2)$. Thus, $(0,0)$ denotes the lowest state of the transformed internal problem, while $(1,0)$ and $(0,1)$ correspond to the first excitations of the two auxiliary modes. These labels characterize the states in the two-dimensional Levi-Civita representation and should not be identified directly with the complete three-dimensional quarkonium spectroscopic quantum numbers $(n,L,S,J)$.

We now define the star creation and annihilation operators:
\begin{eqnarray}
	\widehat{a} & = &\left(\sqrt{\frac{m\omega}{2}}q_1\star+i\sqrt{\frac{1}{2m\omega}}p_1\star\right),
	\nonumber\\
	\widehat{a}^\dagger & = & \left(\sqrt{\frac{m\omega}{2}}q_1\star-i\sqrt{\frac{1}{2m\omega}}p_1\star\right),
	\nonumber\\
	\widehat{b} & = & \left(\sqrt{\frac{m\omega}{2}}q_2\star+i\sqrt{\frac{1}{2m\omega}}p_2\star\right),
	\nonumber\\
	\widehat{b}^\dagger & = & \left(\sqrt{\frac{m\omega}{2}}q_2\star-i\sqrt{\frac{1}{2m\omega}}p_2\star\right),
	\label{b18}
\end{eqnarray}

These operators satisfy the following relations:
\begin{eqnarray}
 & & [\widehat{a},\widehat{a}^{\dagger}]=[\widehat{b},\widehat{b}^{\dagger}]=1,\nonumber\\
& & \widehat{a}\,\phi_{n_1}(q_1,p_1)=\sqrt{n_1}\,\phi_{n_1-1}(q_1,p_1),\nonumber\\
	& & \widehat{a}^\dagger\,\phi_{n_1}(q_1,p_1)=\sqrt{n_1+1}\,\phi_{n_1+1}(q_1,p_1),\nonumber\\
 & & \widehat{b}\,\phi_{n_2}(q_2,p_2)=\sqrt{n_2}\,\phi_{n_2-1}(q_2,p_2),\nonumber\\
	& & \widehat{b}^\dagger\,\phi_{n_2}(q_2,p_2)=\sqrt{n_2+1}\,\phi_{n_2+1}(q_2,p_2).
\label{b29}
\end{eqnarray}
In addition, it is useful to note that  
\begin{eqnarray}
q_{k}\star & = & \frac{1}{\sqrt{2m\omega}}\left ( \widehat{a} + \widehat{a}^{\dagger} \right ),
\nonumber\\
p_{k}\star & = & -i \sqrt{\frac{m\omega}{2}}\left ( \widehat{a}-\widehat{a}^{\dagger} \right ),
\end{eqnarray}
$k=1,2$, and
\begin{equation}
\widehat{a}^{\dagger}\widehat{a}\,\psi _{n_{1},n_{2}}^{(0)}=n_{1}\psi_{n_{1},n_{2}}^{(0)},\hspace{3mm}
\widehat{b}^{\dagger}\widehat{b}\,\psi _{n_{1},n_{2}}^{(0)}=n_{2}\psi_{n_{1},n_{2}}^{(0)}.
\end{equation}
The unperturbed and perturbed Hamiltonian are, respectively, rewritten as 
\begin{eqnarray}
\widehat{H}_{0} & = & \omega\left( \widehat{a}^{\dagger}\widehat{a}+\widehat{b}^{\dagger}\widehat{b}+1 \right ),
\nonumber\\
\widehat{H}_{1} & = &
\frac{B^2}{64m^4\omega^3}
\left[(\widehat{a}+\widehat{a}^{\dagger})^{2}+(\widehat{b}+\widehat{b}^{\dagger})^{2}\right]^{3}+\frac{\beta}{m^2\omega^2}\left[(a+a^{\dagger})^{2}+(b+b^{\dagger})^{2}\right]^{2}.
\label{b21}
\end{eqnarray}
The eigenvalues of the undisturbed Hamiltonian in the state $\phi_{n_1,n_2}$ are then
\begin{equation}
\alpha_{n_{1},n_{2}}^{(0)}=\frac{1}{4}(n_{1}+n_{2}+1)\omega.
\label{a}
\end{equation}
As addressed before, in this zero order, the parameter $\alpha_{n_{1},n_{2}}^{(0)}$ has to be restricted to the original value as $\alpha$ in the Cornell potential. Therefore, by using the definitions of $\omega$ and Eq.~\eqref{a}, we obtain the zero order energy spectrum
\begin{equation}
     E_{n_1,n_2}^{(0)}
  =  - \frac{2m\alpha^2}{(\,n_1 + n_2 + 1\,)^2}
  \;\mp\;
  \frac{B}{2m}.\label{energ1}
\end{equation}
It is important to observe that the limit $B\rightarrow 0$ these result agrees with previous zero-field results~\cite{campos2018,Yang}.

The quasi-amplitudes of probability functions are obtained  by noticing that 
\begin{equation}
\widehat{a}\,\phi_{0}=0,\hspace{3mm} \widehat{b}\,\phi_{0}=0.
\end{equation}
Using Eq.~(\ref{b18}), it leads to
\begin{equation}
   \left(\sqrt{\frac{m\omega}{2}}q_1\star+i\sqrt{\frac{1}{2m\omega}}p_1\star\right)\phi_{0}=0,
\end{equation}
such that
\begin{equation}
\label{b30}
\Psi_{n_1,n_2}^{(0)}(q_1,p_1,q_2,p_2)=N e^{-\Big(m\omega q_{1}^2+\frac{p_{1}^2}{m\omega}\Big)}L_{n_1}\Big(2\Big(m\omega q_{1}^2+\frac{p_{1}^2}{m\omega}\Big)\Big)e^{-\Big(m\omega q_{2}^2+\frac{p_{2}^2}{m\omega}\Big)}L_{n_2}\Big(2\Big(m\omega q_{2}^2+\frac{p_{2}^2}{m\omega}\Big)\Big),
\end{equation}
where $L_{n}$ are Laguerre polynomials of order $n$ and $N$ is a normalization constant. The excited states are derived
by using the raising operators $\widehat{a}^\dagger$ and $\widehat{b}^\dagger$ as is usual.

\subsection{First order correction}
The first-order correction to the eigen-functions in usual perturbation theory is given by
\begin{equation}
\label{b32}
\Psi_{n_1,n_2}^{(1)}(q_1,p_1,q_2,p_2)=\Psi_{n_1,n_2}^{(0)}(q_1,p_1,q_2,p_2)
+\sum_{(m_1,m_2)\neq(n_1,n_2)}\frac{{\cal I}^{}}{4\alpha_{n_1,n_2}^{(0)}-4\alpha_{m_1,m_2}^{(0)}} \Psi^{(0)}_{m_1,m_2}(q_1,p_1,q_2,p_2),
\end{equation}
where
\begin{eqnarray}\label{EqIntegral}
	{\cal I}^{} & = &
 \langle \Psi_{m_1,m_2}^{\ast(0)}(q_1,p_1,q_2,p_2)
	|\widehat H_1|
\Psi_{n_1,n_2}^{(0)}(q_1,p_1,q_2,p_2)\rangle.
\end{eqnarray}
Using the orthogonality relations
\begin{equation}\label{ortob1}
 \langle\phi_{n}^{\ast}(q_1,p_1)|\phi_{m}(q_1,p_1)\rangle=\delta_{n,m},
\end{equation}
\begin{equation}\label{ortob2}
 \langle\Gamma_{n}^{\ast}(q_2,p_2)|\Gamma_{m}(q_2,p_2)\rangle=\delta_{n,m},
\end{equation}
 the meson  ground state is then (details in the \cref{app-B})
 
\begin{align}
\nonumber \Psi_{0,0}^{(1)} &= \Psi_{0,0}^{(0)}
-\frac{1}{\omega} \Bigg[
\left( \frac{9\sqrt{2}\,B^2}{16m^4\omega^3}
+ \frac{4\sqrt{2}\,\beta}{m^2\omega^2}
\right)\Psi_{2,0}^{(0)}+ \left( \frac{9B^2}{32m^4\omega^3}
+ \frac{\beta}{m^2\omega^2} \right)\Psi_{2,2}^{(0)}
+ \left(\frac{9\sqrt{6}\,B^2}{64m^4\omega^3}
+ \frac{\sqrt{6}\,\beta}{2m^2\omega^2} \right)\Psi_{4,0}^{(0)}
\\
&+ \frac{\sqrt{3}\,B^2}{32m^4\omega^3}
\Psi_{4,2}^{(0)} +\frac{\sqrt{5}\,B^2}{32m^4\omega^3}\Psi_{6,0}^{(0)}\Bigg].
\label{b40}
\end{align}

The eigenfunction for the first excited state is given by
\begin{align}
\nonumber
\Psi_{1,0}^{(1)}
={}& \Psi_{1,0}^{(0)}
-\frac{1}{\omega}\Bigg[\sqrt{2}
\left( \frac{9B^2}{8m^4\omega^3} + \frac{6\beta}{m^2\omega^2}\right) \Psi_{1,2}^{(0)}
+
\sqrt{6}
\left(
\frac{3B^2}{16m^4\omega^3}
+
\frac{\beta}{2m^2\omega^2}
\right)
\Psi_{1,4}^{(0)}
+
\sqrt{6}
\left(
\frac{9B^2}{8m^4\omega^3}
+
\frac{6\beta}{m^2\omega^2}
\right)
\Psi_{3,0}^{(0)}
\\
\nonumber
&+
\sqrt{3}
\left(
\frac{3B^2}{8m^4\omega^3} + \frac{\beta}{m^2\omega^2} \right)
\Psi_{3,2}^{(0)} + \frac{3B^2}{32m^4\omega^3} \Psi_{3,4}^{(0)} + \frac{\sqrt{5}B^2}{32m^4\omega^3} \Psi_{1,6}^{(0)}+ \sqrt{30} \left( \frac{3B^2}{16m^4\omega^3}
+
\frac{\beta}{2m^2\omega^2}
\right)
\Psi_{5,0}^{(0)}
+ \\
&\frac{\sqrt{15}B^2}{32m^4\omega^3}
\Psi_{5,2}^{(0)}
 +\frac{\sqrt{35}B^2}{32m^4\omega^3}
\Psi_{7,0}^{(0)}
\Bigg].
\label{b41b}
\end{align}
and
\begin{align}
\nonumber
\Psi_{0,1}^{(1)}
={}&
\Psi_{0,1}^{(0)} -\frac{1}{\omega}\Bigg[\sqrt{2}\left( \frac{9B^2}{8m^4\omega^3}
+
\frac{6\beta}{m^2\omega^2}\right) \Psi_{2,1}^{(0)}
+
\sqrt{6}\left(\frac{9B^2}{8m^4\omega^3} + \frac{6\beta}{m^2\omega^2}\right)
\Psi_{0,3}^{(0)}
+
\sqrt{3}\left(\frac{3B^2}{8m^4\omega^3}
+
\frac{\beta}{m^2\omega^2}\right)\Psi_{2,3}^{(0)}
\\
\nonumber
&+
\sqrt{6}
\left(
\frac{3B^2}{16m^4\omega^3}
+
\frac{\beta}{2m^2\omega^2}
\right)
\Psi_{4,1}^{(0)}
+
\sqrt{30}
\left(
\frac{3B^2}{16m^4\omega^3}
+
\frac{\beta}{2m^2\omega^2}
\right)
\Psi_{0,5}^{(0)}
\\
\nonumber
&+
\frac{3B^2}{32m^4\omega^3}
\Psi_{4,3}^{(0)}
+
\frac{\sqrt{5}B^2}{32m^4\omega^3}
\Psi_{6,1}^{(0)}
\\
&+
\frac{\sqrt{15}B^2}{32m^4\omega^3}
\Psi_{2,5}^{(0)}
+
\frac{\sqrt{35}B^2}{32m^4\omega^3}
\Psi_{0,7}^{(0)}
\Bigg].
\label{eq:Psi01}
\end{align}
The Wigner function for quark-antiquark bound states is then obtained from these eigen-states by using the $\star$ product as
\begin{equation}
\label{wigb}
f_W(q_1,p_1,q_2,p_2)=\Psi_{n_1,n_2}^{(1)}(q_1,p_1,q_2,p_2)\star \Psi_{n_1,n_2}^{\dagger(1)}(q_1,p_1,q_2,p_2). 
\end{equation}
 Since the quasi-amplitudes considered here are real,
Eq.~\eqref{wigb} becomes
\begin{equation}
f_W
=
f_W^{(0)}
+
2\lambda\,\mathrm{Re}
\left[
\Psi_{0,0}^{(0)}
\star
\delta\Psi_{0,0}^{(1)}
\right]
+
\mathcal{O}(\lambda^2).
\label{eq:wigner_ground_first_order_compact}
\end{equation}
where $\lambda$ is a bookkeeping perturbation parameter, set to unity at the end of the calculation.

The first-order correction for the eigenvalues is obtained from
\begin{equation}
\label{energia1}
4\Delta \alpha^{(1)}_{n_1,n_2}=\int \Psi_{n_1,n_2}^{(0)}\widehat{H}_1\Psi_{n_1,n_2}^{\dagger(0)}dq_1dp_1dq_2dp_2,
\end{equation} 
Substituting $\widehat{H}_1$, we have
\begin{align}
4\Delta\alpha_{n_1,n_2}^{(1)}
={}&
\int
\Psi_{n_1,n_2}^{(0)*}
\Bigg\{
\frac{B^2}{64m^4\omega^3}
\left[
(\hat a+\hat a^\dagger)^2
+
(\hat b+\hat b^\dagger)^2
\right]^3
\nonumber\\
&\qquad\qquad
+
\frac{\beta}{m^2\omega^2}
\left[
(\hat a+\hat a^\dagger)^2
+
(\hat b+\hat b^\dagger)^2
\right]^2
\Bigg\}
\Psi_{n_1,n_2}^{(0)}
\,dq_1\,dp_1\,dq_2\,dp_2 .
\end{align}
which yields 
\begin{equation}
\Delta\alpha_{n_1,n_2}^{(1)}
=
\frac{B^2}{256m^4\omega^3}\Delta'
+
\frac{\beta}{4m^2\omega^2}\Delta''.
\label{delta}
\end{equation}
where
\begin{align}
\Delta'
={}&
20(n_1^3+n_2^3)
+ 36(n_1^2n_2+n_1n_2^2) +48(n_1^2+n_2^2) + 72n_1n_2 + 76(n_1+n_2) + 48.
\label{eq:Delta_prime}
\end{align}
and
\begin{align}
    \Delta''=\Big[
6n_1^2+6n_2^2+8n_1n_2
+10n_1+10n_2+8
\Big].
\end{align}

In this way, we obtain the result for the first-order corrected eigenvalue
\begin{equation}
\alpha_{n_1,n_2}^{(1)}
=
\frac{1}{4}(n_1+n_2+1)\omega
+
\frac{B^2}{256m^4\omega^3}\Delta'
+
\frac{\beta}{4m^2\omega^2}\Delta''.
\end{equation}

For the ground state ($n_{1}=0,n_{2}=0$), the second-order perturbation eigenvalue correction is given by
\begin{align}
4\alpha_{0,0}^{(2)} \simeq{}&\omega
+ \left( \frac{3B^2}{4m^4\omega^3} +\frac{8\beta}{m^2\omega^2}\right)
+\left(-\frac{249B^4}{64m^8\omega^7}
-\frac{45B^2\beta}{m^6\omega^6}
-\frac{144\beta^2}{m^4\omega^5}\right).
\label{eq:alpha}
\end{align}
Using the Eq.~\eqref{energ1} and Eq.~\eqref{delta}, the energy spectrum up to first order is calculated as
\begin{equation}
E_{n_1,n_2}^{(1)}
=
-\frac{m}{8}
\left[
\frac{
4\alpha
-
\dfrac{B^2}{64m^4\omega^3}\Delta'
-
\dfrac{\beta}{m^2\omega^2}\Delta''
}{
n_1+n_2+1
}
\right]^2
\mp\frac{B}{2m}.
\label{enefim}
\end{equation}
This result gives the energy eigenvalues for a quarkonium particle described by the Cornell potential in the presence of an external field.
Specific cases are analyzed in the next section.

\section{Stationary-state Wigner functions }
\label{dr}
In this section we analyze the Wigner function of the stationary states describing  the quark-antiquark system, up to the first order
approximation,  considering the $c\overline{c}$, $b\overline{b}$, mesons obtained by using the symplectic quantum mechanics approach.
In the lowest bound state $c\overline{c}$ has a mass of roughly $3.5$ times greater
than the proton mass~\cite{Godfrey,Mansour,RL_1,renato2024frac}. We obtained here
the ground state with the respective quasi-amplitude probability. For the numerical analysis, 
 the following parameter values are used~\cite{Mutuk2019}: 
the confinement parameter $\beta=0.191\,{\rm GeV}^{2}$, charm quark mass $m_{c}=1.3205\,{\rm GeV}$, bottom quark mass $m_{b}=4.7485\,{\rm GeV}$,
the reduced mass $m_{c\overline{c}}=0.6602\,{\rm GeV}$, $m_{b\overline{b}}=2.374\,{\rm GeV}$. For the numerical phase-space analysis, the frequency values $\omega=1.9658$ for $c\bar c$ and $\omega=1.7977$ for $b\bar b$
are used in the graphical representation of the Wigner functions. In addition, we also take the parameters $q_1=q$ and $p_1=p$ with ($q_2=p_2=0$) fixed in the graphical representation. 
In the phase-space cuts used below, $q_2=0$ and $q_1=q$. Therefore, from the Levi-Civita relation $r=q_1^2+q_2^2$, one has $r=q^2$. Hence, the horizontal variable $q^2$, with units of ${\rm GeV}^{-1}$, represents the physical inter-quark distance in the chosen phase-space cut.

\begin{figure}[H]
\centering
\includegraphics[scale=0.6]{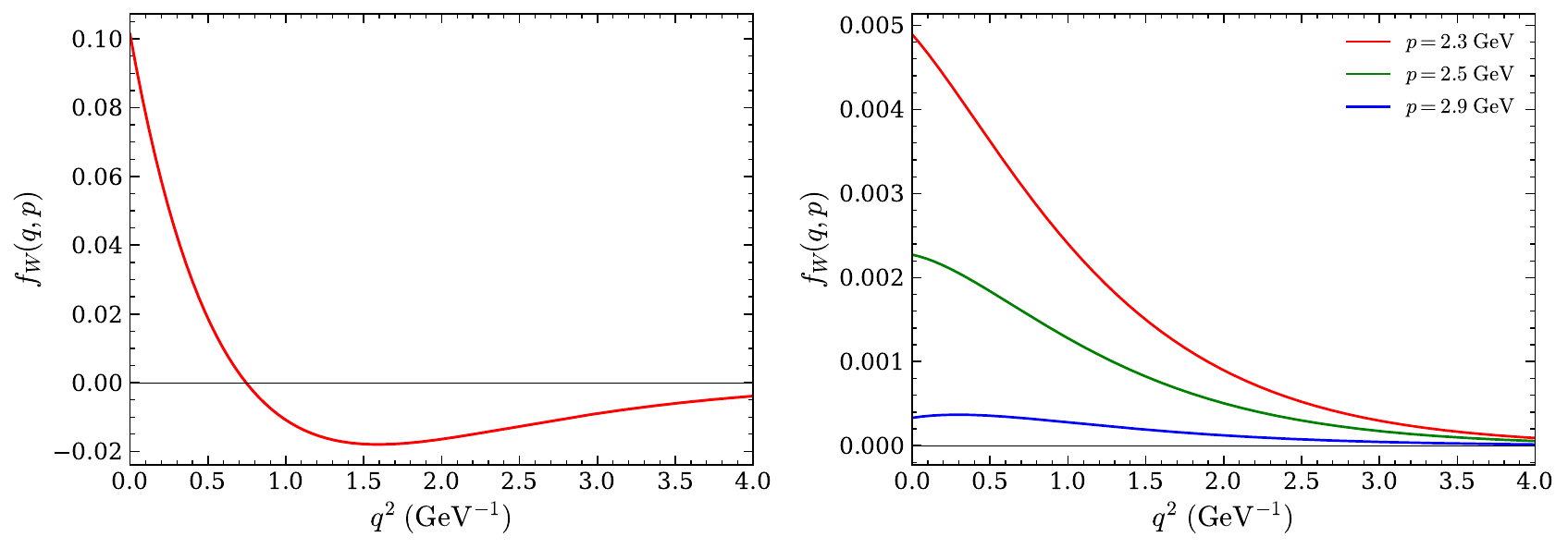} 
\caption{ The left graph, first-order corrected Wigner function for the ground state ($n_{1}=0$, $n_{2}=0$) of the $c\overline{c}$ meson, with momentum $p=0\,\text{GeV}$ and no magnetic field ($B=0\,\text{GeV}^2$), plotted as a function of the interquark distance $q^{2}$ in the region $0 \leq q^{2} \lesssim 4\,\text{GeV}^{-1}$. The right graph, the Wigner function is presented for the same state with momenta $p=2.3\,\text{GeV}$ (red), $2.5\,\text{GeV}$ (green), and $2.9\,\text{GeV}$ (blue), also for $B=0\,\text{GeV}^2$. }
\label{fig:graph1}
\end{figure}

\begin{figure}[H]
\centering
\includegraphics[scale=0.6]{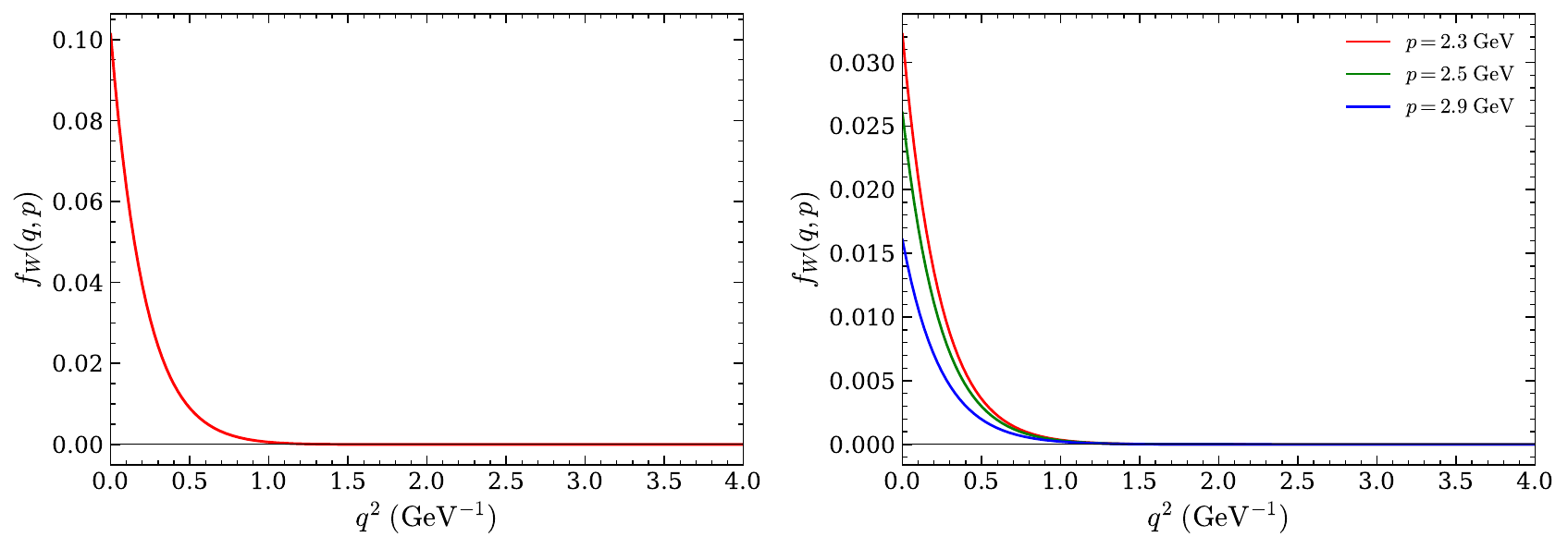} 
\caption{The left graph, Wigner function with first-order correction for the state ($n_{1}=0$, $n_{2}=0$) (ground state) of the $b\overline{b}$ meson with $p=0\,\text{GeV}$ and $B=0\,\text{GeV}^2$. The right graph, the Wigner function is plotted for the same state with momenta $p=2.3\,\text{GeV}$ (red), $2.5\,\text{GeV}$ (green), and $2.9\,\text{GeV}$ (blue), also for $B=0\,\text{GeV}^2$.}
\label{fig:graph2}
\end{figure}

The left graphics of Fig.~(\ref{fig:graph1}) and Fig.~(\ref{fig:graph2}), display the first-order corrected behavior of the Wigner function $f_{W}(q_{1},p_{1};q_{2}=0,p_{2}=0)$ for the ground state ($n_1 = 0, n_2 = 0$) of the mesons $c\bar{c}$ and $b\bar{b}$, with fixed momentum $p = 0\, \,\text{GeV}$ and the right graphics for momentum variations $p=2.3\,\text{GeV}$ (red), $2.5\,\text{GeV}$ (green), and $2.9\,\text{GeV}$ (blue) with no external magnetic field ($B = 0$).

In the graphics on the left of Fig.~(\ref{fig:graph1}) and Fig.~(\ref{fig:graph2}) we observe that for momentum $p=0\,\, {\rm GeV}$, and for $B=0\, \, {\rm GeV^2}$, the Wigner function exhibits the graphical behavior in accordance with a previous result~\cite{RL_1}. For $p=0$, the Wigner function is predominantly localized within the displayed interval $0\leq q^2\lesssim4\,{\rm GeV}^{-1}$. This behavior
characterizes the phase-space localization obtained from the confining Cornell interaction within the present effective description and is consistent with previous results~\cite{RL_1}. In  Fig.~(\ref{fig:graph1}) (right graph) and Fig.~(\ref{fig:graph2}) (right graph), it is observed that by varying  the momentum $p=2.3\,\text{GeV}$ (red line), $p=2.5\,\text{GeV}$ (green line), and $p=2.9\,\text{GeV}$ (blue line), with $B=0\,\text{GeV}^2$, the Wigner function shifts to the left and the curve decrease within the region $q^{2}=0\,$ to $q^{2}\approx 4\,{\rm GeV^{-1}}$.
For larger values of momentum, the Wigner function becomes strongly suppressed within the phase-space region displayed in Fig.~(\ref{fig:graph1}) and Fig.~(\ref{fig:graph2}). This behavior indicates a progressive suppression of the corresponding phase-space distribution (the field describing the system) in the considered region.  
Within the symplectic quantum-mechanics representation, this behavior provides a phase-space signature consistent with the confining structure of the Cornell potential. The symplectic formulation allows this behavior to be analyzed directly through the geometry of phase space, which is no longer obvious if the started point is the Euclidian representation. These findings, consistent with previous results~\cite{RL_1,rluz2023frac}, are associated with the confined region of a hadron.

\begin{figure}[H]
\centering
\includegraphics[scale=0.6]{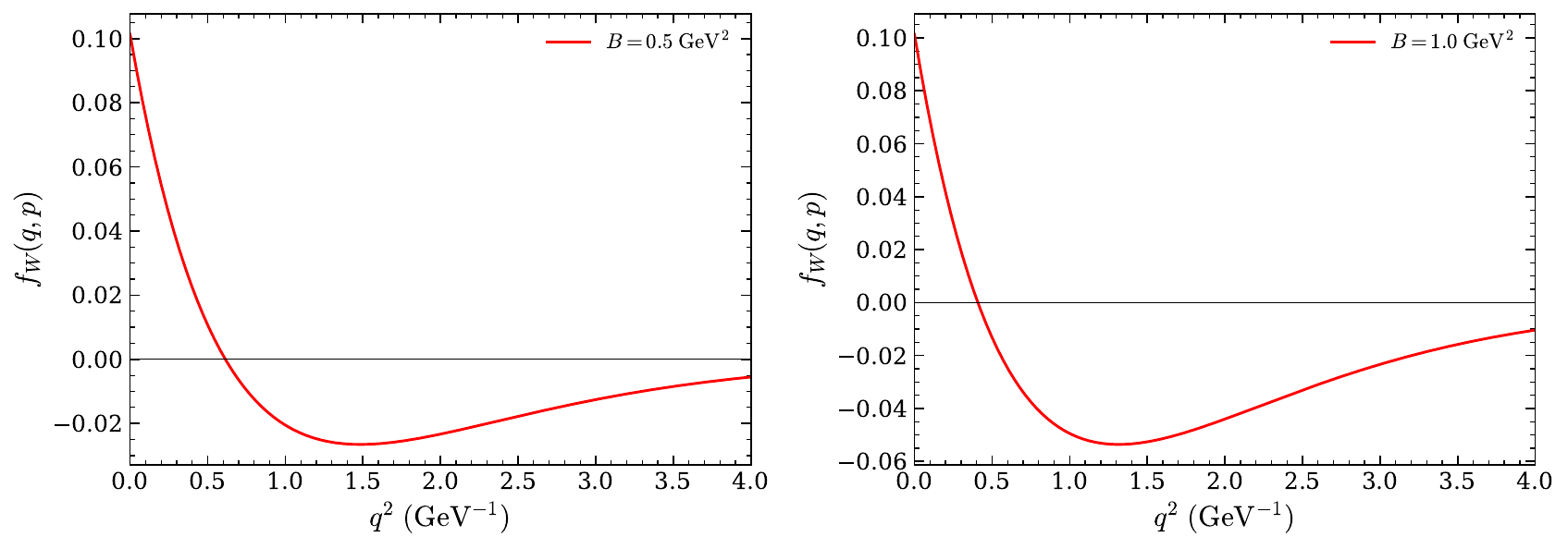} 
\caption{Wigner function including the first-order correction for the ground state 
$(n_1=0,n_2=0)$ of the $c\bar{c}$ meson, plotted as a function of $q^2$ for $p=0$ GeV. The left and right graphics correspond to magnetic fields $B=0.5$ GeV$^2$ and $B=1.0$ GeV$^2$, respectively.}
\label{fig:mesonccB}
\end{figure}
\begin{figure}[H]
\centering
\includegraphics[scale=0.6]{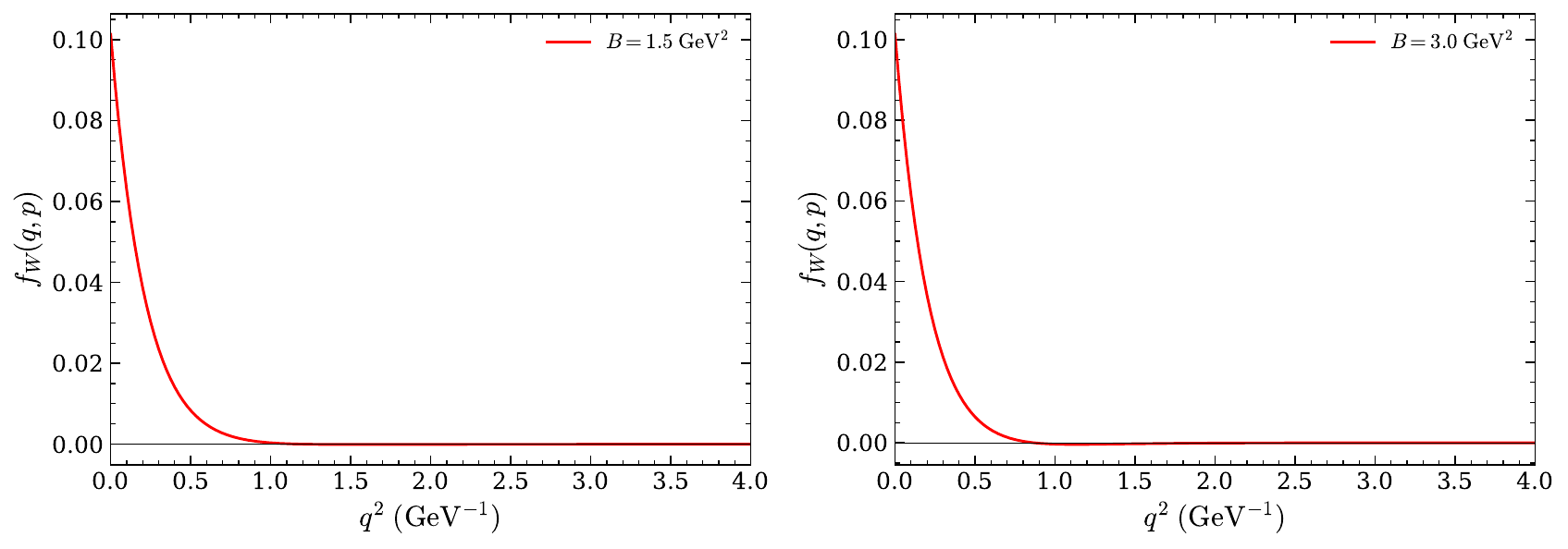} 
\caption{Wigner function with first-order correction for the ground state ($n_{1}=0$, $n_{2}=0$) of the $b\overline{b}$ meson with $p=0\,\text{GeV}$. The graph on the left presents the results for a magnetic field $B=1.5\,\text{GeV}^2$, while the graph on the right considers the $B=3.0\,\text{GeV}^2$. }
\label{fig:mesonbbB}
\end{figure}
Figures~(\ref{fig:mesonccB}) and~(\ref{fig:mesonbbB}) illustrate the first-order corrected Wigner function for the same eigen-state ($n_{1}=0,n_{2}=0$), but considering a non-zero magnetic field. In Fig.~\ref{fig:mesonccB}, the behavior of the Wigner function for the $c\bar{c}$ meson is shown for
$B=0.5\,\text{GeV}^2$ and $B=1.0\,\text{GeV}^2$. In Fig.~\ref{fig:mesonbbB}, the corresponding behavior for the
$b\bar{b}$ meson is shown for $B=1.5\,\text{GeV}^2$ and $B=3.0\,\text{GeV}^2$. Note that, as the magnetic field changes, the Wigner functions develop different negative regions. This negativity is an indicator of the non-classicality of the system. The field values used in the graphical analysis are illustrative values
chosen within the perturbative regime discussed below.

To assess the perturbative regime, we introduce the ratio
\begin{equation}
R(B)=
\frac{\|\delta\Psi^{(1)}\|}
{\|\Psi^{(0)}\|}
=
\left[
\sum_{(m_1,m_2)\neq(0,0)}
|A_{m_1,m_2;0,0}|^2
\right]^{1/2},
\end{equation}
where the orthogonality of the unperturbed states has been used. This ratio is employed here as a practical self-consistency measure of the first-order approximation. For the $c\bar c$ ground state, we obtain $R(0)=0.339$, $R(0.5)=0.417$, and $R(1.0)=0.656$, whereas $R(1.5)=1.057$. Accordingly, the graphical analysis of the
$c\bar c$ ground state is restricted to $B\leq1.0~{\rm GeV}^2$, for which the first-order correction remains smaller than the zeroth-order contribution. For the $b\bar b$ ground state, the corresponding ratios are much smaller, $R(1.5)=0.040$ and $R(3.0)=0.059$, so that the field values employed in Fig.~\ref{fig:mesonbbB} remain well within the perturbative regime. Next, the nonclassical character of the states is quantified through the Wigner-negativity parameter. Finally, the mass spectrum is analyzed, and the experimental ground-state masses are used to constrain the magnetic-field value $B$. In the following the statistical nature of the Wigner function is analyzed. 

The negativity of the Wigner function provides a quantitative indicator of the nonclassical statistical character of a quantum state. This measure is based on the total volume of the negative regions of the
Wigner function in phase space. For a normalized Wigner function, a purely positive distribution gives a vanishing negativity parameter, whereas negative contributions lead to a nonzero value. In the present case, for the quark-antiquark bound state described by the Wigner function
$f_W(q_1,p_1,q_2,p_2)$, the negativity parameter is defined as~\cite{Kenfack2004}
\begin{align}
\eta(\Psi)&=\int_{}\left|f_W(q_1,p_1,q_2,p_2)\right|\,dq_1\,dp_1\,dq_2\,dp_2-1 \nonumber\\ 
&=
2
\int_{f_W<0}
\left|
f_W(q_1,p_1,q_2,p_2)
\right|
\,dq_1\,dp_1\,dq_2\,dp_2 .
\label{eq:kz_negativity}
\end{align}
where $\int_{}f_W(q_1,p_1,q_2,p_2)\,dq_1\,dp_1\,dq_2\,dp_2=1 $. This indicator represents the doubled volume of the integrated part of the Wigner function. Next, we numerically calculate this parameter for the $c\overline{c}$ ground state at different values of $B$. The perturbative behavior of the first excited states is discussed separately below.
\begin{table}[H]
\centering
\caption{Wigner-negativity parameter $\eta(\Psi)$ and perturbative
ratio $R(B)$ for the ground state of the $c\bar c$ meson at different
magnetic-field strengths.}
\label{tab:negativity_ccbar}

\begin{tabular}{c c c c c}
\hline\hline
$n_1,n_2$ & State & $B~({\rm GeV}^2)$ & $R(B)$ & $\eta(\Psi)$\\
\hline
$0,0$ & $\Psi_{0,0}^{(1)}$ & $0$   & $0.3388$ & $0.2146$\\
$0,0$ & $\Psi_{0,0}^{(1)}$ & $0.5$ & $0.4174$ & $0.3267$\\
$0,0$ & $\Psi_{0,0}^{(1)}$ & $1.0$ & $0.6562$ & $0.7079$\\
\hline\hline
\end{tabular}
\end{table} 

\begin{table}[H]
\centering
\caption{Negativity parameter $\eta(\Psi)$ and perturbative ratio $R(B)$ for the ground and first excited states of the $b\bar b$ meson under different magnetic-field strengths.}
\label{tab:negativity_bb}

\begin{tabular}{c c c c c}
\hline\hline
$n_1,n_2$ & State & $B~({\rm GeV}^2)$ & $R(B)$ & $\eta(\Psi)$\\
\hline

$0,0$ & $\Psi_{0,0}^{(1)}$ & $0$
& $0.0343$ & $0.00016$\\

$1,0$ & $\Psi_{1,0}^{(1)}$ & $0$
& $0.1010$ & $0.43374$\\

$0,1$ & $\Psi_{0,1}^{(1)}$ & $0$
& $0.1010$ & $0.43374$\\

$0,0$ & $\Psi_{0,0}^{(1)}$ & $1.5$
& $0.0403$ & $0.00054$\\

$1,0$ & $\Psi_{1,0}^{(1)}$ & $1.5$
& $0.1240$ & $0.44425$\\

$0,1$ & $\Psi_{0,1}^{(1)}$ & $1.5$
& $0.1240$ & $0.44425$\\

$0,0$ & $\Psi_{0,0}^{(1)}$ & $3.0$
& $0.0586$ & $0.00347$\\

$1,0$ & $\Psi_{1,0}^{(1)}$ & $3.0$
& $0.1936$ & $0.50015$\\

$0,1$ & $\Psi_{0,1}^{(1)}$ & $3.0$
& $0.1936$ & $0.50015$\\

\hline\hline
\end{tabular}
\end{table}

In Table~\ref{tab:negativity_ccbar}, the Wigner negativity of the $c\bar c$ ground state increases from $\eta=0.2146$ at $B=0$ to $\eta=0.3267$ at $B=0.5~{\rm GeV}^2$ and $\eta=0.7079$ at $B=1.0~{\rm GeV}^2$. Over this field range, the perturbative ratio remains below unity, with $R=0.3388$, $0.4174$, and $0.6562$,
respectively. Therefore, the increase of the Wigner negativity for the $c\bar c$ ground state is obtained within the perturbatively controlled regime adopted in the present analysis.

For the first excited $c\bar c$ states, the perturbative ratio is already close to unity at zero field, $R\simeq0.999$, and exceeds unity for the nonzero fields considered here. Consequently, although
their first-order Wigner functions and negativity parameters can be formally evaluated, these finite-field values are not used for quantitative conclusions.

In Table~\ref{tab:negativity_bb}, the $b\bar b$ system displays a considerably smaller perturbative correction. For the ground state, $R$ increases from $0.0343$ at $B=0$ to only $0.0586$ at $B=3.0~{\rm GeV}^2$, while for the first excited states the largest value is $R=0.1936$. Thus, all the $b\bar b$ states considered remain
within the perturbative regime. The corresponding Wigner negativity increases from $\eta=0.00016$ to $\eta=0.00347$ for the ground state and from $\eta=0.43374$ to $\eta=0.50015$ for the first excited
states as the magnetic field increases from $B=0$ to $B=3.0~{\rm GeV}^2$.

These results show that the external magnetic field enhances the negative phase-space regions of the Wigner function within the perturbatively controlled cases considered here, providing a quantitative signature of increasing non-classicality. 

The numerical evaluation of the Wigner-negativity parameter was performed on a uniform four-dimensional phase-space grid. We introduce the dimensionless variables $Q=\sqrt{m\omega}\,q$ and $P=p/\sqrt{m\omega}$, with
$d\Gamma=dq_1\,dp_1\,dq_2\,dp_2$, and use the integration domain $|Q_i|\leq Q_{\max}$ and $|P_i|\leq P_{\max}$, $i=1,2$. The final calculations use $Q_{\max}=P_{\max}=6$ and $N=60$ grid points along each phase-space coordinate, corresponding to $N^4=12\,960\,000$ points in the four-dimensional integration grid.

For the $c\bar c$ system, the numerical convergence was explicitly tested for the ground state at $B=1.0~{\rm GeV}^2$, which corresponds to the largest magnetic field retained in the perturbatively controlled charmonium analysis. At fixed $Q_{\max}=P_{\max}=6$, the negativity changes from $\eta=0.710311$ for $N=36$ to $0.707465$, $0.707801$, and $0.707884$ for $N=44$, $52$, and $60$, respectively. The relative difference between the two finest grids, $N=52$ and $N=60$, is approximately $0.012\%$.

The convergence with respect to the integration domain for the $c\bar c$ system was examined while keeping approximately the same grid spacing. For $(L,N)=(5,50)$, $(6,60)$, and $(7,70)$, where $Q_{\max}=P_{\max}=L$, the corresponding values are $\eta=0.707830$, $0.707884$, and $0.707921$, respectively. The total variation over this range is approximately $0.013\%$.

For the $b\bar b$ system, the convergence was tested for the first excited state $(n_1,n_2)=(1,0)$ at
$B=3.0~{\rm GeV}^2$, which corresponds to the largest negativity considered within the perturbatively controlled bottomonium analysis. At fixed $Q_{\max}=P_{\max}=6$, the values $\eta=0.499073$, $0.500147$, $0.502265$, and $0.499162$ were obtained for $N=56$, $60$, $64$, and $68$, respectively. Although a small oscillatory dependence on the grid resolution is observed, the values remain stable around $\eta\simeq0.500$, with a total spread below $0.7\%$ over the range $N=56$--$68$.

The integration-domain convergence for the $b\bar b$ system was also tested at approximately fixed grid spacing. For $(L,N)=(5,50)$, $(6,60)$, and $(7,70)$, the corresponding values are $\eta=0.500104$, $0.500147$, and $0.500217$, respectively, corresponding to a total variation of approximately $0.023\%$.

In all convergence tests, the numerical normalization remains unity within the reported precision. Moreover, the two equivalent numerical expressions $\eta=\int |f_W|\,d\Gamma-1$ and $\eta=2\int_{f_W<0}|f_W|\,d\Gamma$ agree to approximately machine precision. These results confirm that the adopted integration domain and grid resolution provide numerically stable values of the Wigner-negativity parameter.
\begin{table}[H]
\centering
\caption{
Grid-resolution convergence of the Wigner-negativity parameter for representative $c\bar c$ and $b\bar b$ states at $Q_{\max}=P_{\max}=6$.}
\label{tab:conv_negativity}

\begin{tabular}{c c c c c}
\hline\hline
System & State & $B~({\rm GeV}^2)$ & $N$ & $\eta$\\
\hline

$c\bar c$ & $(0,0)$ & $1.0$ & $36$ & $0.710311$\\
$c\bar c$ & $(0,0)$ & $1.0$ & $44$ & $0.707465$\\
$c\bar c$ & $(0,0)$ & $1.0$ & $52$ & $0.707801$\\
$c\bar c$ & $(0,0)$ & $1.0$ & $60$ & $0.707884$\\

\hline

$b\bar b$ & $(1,0)$ & $3.0$ & $56$ & $0.499073$\\
$b\bar b$ & $(1,0)$ & $3.0$ & $60$ & $0.500147$\\
$b\bar b$ & $(1,0)$ & $3.0$ & $64$ & $0.502265$\\
$b\bar b$ & $(1,0)$ & $3.0$ & $68$ & $0.499162$\\

\hline\hline
\end{tabular}
\end{table}

Table~\ref{tab:ccbar_mass_spectrumI} presents the ground-state mass spectrum of the $c\bar c$ and $b\bar b$ mesons obtained within the perturbative treatment developed above, in comparison with experimental data. Table~\ref{tab:ccbar_mass_spectrumIII} provides various theoretical predictions available in the literature which are compared to the experimental results.

The mass spectra is given by the following equation~\cite{Mansour}
\begin{equation}
    M = m_{q}+m_{\overline{q}} - E+V_{0} \label{spectro}.
\end{equation}
Substituting  Eq.~\eqref{frequen} in Eq.~\eqref{spectro}, it leads to 
\begin{equation}
    M = m_{q}+m_{\overline{q}} + V_0 + \frac{m}{8}\omega^{2}
+\frac{(\pm B)}{2m}.
 \label{spectroI}
\end{equation}
where $m_q$ and $m_{\bar q}$ are the masses of the quark and antiquark, respectively, and $m$ is the reduced mass. The frequency and magnetic-field values used in the mass-spectrum analysis are determined simultaneously from the coupled equations discussed in~\cref{app-A}. Table~III shows the mass spectrum corresponding to the ground state of the $c\bar c$ and $b\bar b$ mesons with regard to the experimental data.
\begin{table}[H]
\centering
\caption{Fundamental energy level and mass spectrum of the $c\bar{c}$, $b\bar{b}$ mesons (in GeV), calculated by using the present model and compared with experimental data and various theoretical works. Parameters used: $\alpha = 0.472$, $\beta = 0.191\ \text{GeV}^2$, $V_0= -1.55\ \text{GeV}$ for ($b\bar{b}$), $V_0=0\ \text{GeV}$ for ($c\bar{c}$), $m_c = 1.3205\ \text{GeV}$, $m_{c\overline{c}} = 0.6602\ \text{GeV}$, $m_b = 4.7485\ \text{GeV}$, $m_{b\overline{b}} = 2.374\ \text{GeV}$~\cite{Mutuk2019}. In both of the cases we obtained a value of the external field $B$, that fits the experimental value of $M$.}
\vspace{0.3em}
\begin{tabular}{ c c c c c c c c }
\hline \hline
Meson & State & $\omega$ \,\, & $-B$ $\rightarrow$ & $M$ & $+B$ $\rightarrow$ & $M$ &\,\, Exp.~\cite{pdg} \\
\hline
$c\bar{c}$ & $n_{1}=n_{2}=0$ & \,\, 1.9658 & 0.1810,\, & 2.823 & 0.1810 & 3.097 & 3.097\\
$b\bar{b}$ & $n_{1}=n_{2}=0$ & \,\, 1.7977& 2.6304 & 8.352 & 2.6304 & 9.460 & 9.460 \\
\hline \hline
\end{tabular}
\label{tab:ccbar_mass_spectrumI}
\end{table}

\begin{table}[H]
\centering
\caption{Fundamental energy level and mass spectrum of the $c\bar{c}$, $b\bar{b}$ mesons (in GeV), calculated by using the different model and different parameter choices~\cite{Mutuk2019}.}
\vspace{0.3em}
\begin{tabular}{c c c c c c c c c c c c}
\hline \hline
Meson& $M \rightarrow$& Ref.~\cite{AbuSh2023} & Ref.~\cite{Kumar} & Ref.~\cite{Omugbe1} & Ref.~\cite{rluz2023frac} & Ref.~\cite{renato2024frac} & Ref.~\cite{Abdel-Karim} & Ref.~\cite{Abdel}  &Ref.~\cite{Mutuk2019} & Exp.~\cite{pdg} \\
\hline
$c\bar{c}$ & & 3.074 & 3.078 & 3.098 & 3.1003  & 3.0966 & 3.0954 & 3.095&3.098 & 3.097 \\
$b\bar{b}$ && 9.460 & 9.506 & 9.681 & 9.4818 & 9.4564 & 9.74473 & 9.460&9.460 & 9.460 
\\
\hline \hline
\end{tabular}
\label{tab:ccbar_mass_spectrumIII}
\end{table}
In Table~\ref{tab:ccbar_mass_spectrumI}, for the $c\overline{c}$ meson ground state at $\omega=1.9658\,$, the mass spectrum equation exhibits the two field orientations associated with the magnetic-field term in Eq.~\eqref{spectroI}. Specifically, the model gives $B=0.1810\,{\rm GeV^2}$ for $M=3.097\,{\rm GeV}$, in agreement with the experimental value, while the opposite field orientation, $B=-0.1810\,{\rm GeV^2}$, yields $M=2.823\,{\rm GeV}$. When the field value is taken to be positive ($+B$), the mass spectrum of the $c\bar{c}$ meson is consistent with the experimental value $3.097\,{\rm GeV}$~\cite{pdg}. For the opposite orientation, the predicted mass differs from the experimental ground-state value.

In Table~\ref{tab:ccbar_mass_spectrumI}, for the $b\overline{b}$ meson ground state at $\omega=1.7977\,$, the spectrum also exhibits the two field orientations associated with the magnetic-field term in Eq.~\eqref{spectroI}. Our model gives $B=2.6304\,{\rm GeV^2}$ for $M=9.460\,{\rm GeV}$, while the opposite field orientation, $B=-2.6304\,{\rm GeV^2}$, yields $M=8.352\,{\rm GeV}$. In the case of the $b\bar{b}$ meson, the positive field value $(+B)$ is consistent with the experimental value $M=9.460\,{\rm GeV}$~\cite{pdg}, whereas the opposite orientation leads to a different mass value.

Table~\ref{tab:ccbar_mass_spectrumIII} benchmarks the ground-state masses against several theoretical determinations~\cite{AbuSh2023,Kumar,Mutuk2019,rluz2023frac,Abdel-Karim,renato2024frac,Omugbe1,Abdel} and experiment~\cite{pdg}. For the $c\overline{c}$ meson, the values are near the experimental mass $3.097\,{\rm GeV}$, spanning approximately $3.074-3.1003\,{\rm GeV}$, while for the $b\overline{b}$ meson they span roughly $9.4564-9.74473\,{\rm GeV}$, compared with the experimental value $9.460\,{\rm GeV}$. Note that this comparison indicates that the ground-state predictions are of the same order and precision as widely used approaches in the literature, with the present results reproducing the corresponding experimental ground-state masses as shown in Table~\ref{tab:ccbar_mass_spectrumI}.

Although the perturbative expressions derived here are valid for arbitrary $(n_1,n_2)$, in this work we present explicit numerical results mainly for the ground-state $1S$ heavy quarkonia, since the ground state provides the most accurately measured experimental data. Regarding the quantitative results of the model presented here, the values obtained for the parameter $B$, given in $\text{GeV}^2$, correspond to very intense effective magnetic fields. Such field strengths are relevant in the context of strong-interaction systems, where transient magnetic fields produced in non-central heavy-ion collisions can reach extremely large values at RHIC and LHC~\cite{Delia2011}. In the present approach, $B$ should therefore be understood as the external magnetic-field parameter entering the effective quarkonium model.

To assess the stability of the magnetic-field parameter, we performed
a local sensitivity analysis for the $c\bar c$ ground state, as shown
in Table~V (details in the \cref{app-C}). Since not all model parameters
have quoted statistical uncertainties, these values should be interpreted
as sensitivity estimates rather than experimental error bars.

\begin{table}[H]
\centering
\caption{Local sensitivity analysis of the magnetic field $B$ for the $c\bar c$ ground state. The central values are $\omega_0=1.9658$ and $B_0=0.1810~{\rm GeV}^2$}. A relative variation of $1\%$ was used for each nonzero input parameter, while $\Delta V_0=0.01~{\rm GeV}$ was used since $V_0=0$ for the $c\bar c$ system. The quantity $|\Delta B_i|$ represents the change in the magnetic field caused by the local variation of each input parameter.
\label{tab:sensitivity_ccbar}

\begin{tabular}{c c c c c}
\hline\hline
Parameter $p_i$ & Central value & $\Delta p_i$
& $\partial B/\partial p_i$ & $|\Delta B_i|$ \\
\hline

$M$
& $3.097$ & $0.03097$ & $+1.224$ & $0.0379$
\\
$m_c=m_{\bar c}$
& $1.3205$ & $0.01321$ & $-2.447$ & $0.0323$
\\
$V_0$
& $0$ & $0.010$ & $-1.224$ & $0.0122$
\\
$m$
& $0.6602$ & $0.00660$ & $+0.177$ & $0.00117$
\\
$\alpha$
& $0.472$ & $0.00472$ & $-0.604$ & $0.00285$
\\
$\beta$
& $0.191$ & $0.00191$ & $-0.818$ & $0.00156$
\\
\hline\hline
\end{tabular}
\end{table}

The sensitivity analysis shows that the magnetic field is mainly affected by the meson mass and the effective quark masses, while its dependence on $V_0$, $\alpha$, $\beta$, and the reduced mass is comparatively smaller for the chosen local variations.

\section{Final concluding remarks}
\label{fr}
In the present work, the symplectic quantum mechanics has been applied to obtain the ground and excited states of a quark-antiquark system,  within a perturbative framework. The quarks are  interacting through the Cornell potential in  an external magnetic field. 

First, a symplectic representation of the  Pauli-Schrödinger-type equation has been derived by using a phase space representation of Galilei group.   This equation  is then solved by following a Levi-Civita mapping associated with a perturbative method. This theoretical procedure leads to the Wigner function for the quark-antiquark bound state. Aspects of the confinement are then investigated, providing consistent results with  experimental data.  These results are  inherently manifest due to the phase-space geometry. The effect of the external field on the statistical nature of the quarkonium state is analyzed. In this sense, the non-classicality discussed here refers specifically to the Wigner negativity in phase space, quantified numerically through the negativity parameter for the ground state and the first excited meson states at different values of the external magnetic field. Therefore, this result should be interpreted as a phase-space signature of quantum interference.

The experimental spectrum is considered in order to estimate the intensity of the external field, $B$. The results for the mass spectrum are, in particular, compared with other models in the literature. The field obtained for the $c\bar c$ system lies within the strong-field scale relevant to transient magnetic fields produced in non-central heavy-ion collisions. Such transient fields can reach approximately $10^{14}$~T at RHIC and $10^{15}$~T at LHC~\cite{Delia2011}. For the $b\bar b$ system, the present model yields a substantially larger effective field, corresponding to a stronger-field regime than the typical values estimated for RHIC and LHC collisions. Therefore, the comparison with transient heavy-ion magnetic fields should be understood as a comparison of physical scales, rather than as an identification of all the field values obtained here with the magnetic fields produced in these collisions.

Finally, regarding the perturbative approach, the main part of the analysis has been  carried out up to first order approximation, although the energy spectrum has also been derived up to second order. A detailed analysis of higher orders in the perturbative method is an aspect that demands significantly more computational effort; and as such, these results will be presented elsewhere.

\section*{Acknowledgments}

This work is partially supported by the Brazilian funding agency CNPq (Conselho Nacional de Desenvolvimento Científico e Tecnológico-Brazil), grant numberis 312857/2021-7 (TMRF). 

\appendix
\section{}
\label{app-A}
Here, we have shown the explicit calculations of frequency and magnetic field values in agreement with the experimental mass spectrum for the mesons $c\bar{c}$ and $b\bar{b}$. We use the already defined parameters and the respective equations.

For the ground state $(n_1=0,n_2=0)$, the second-order perturbation eigenvalue correction is given by

\begin{equation}
\begin{split}
4\alpha_{0,0}^{(2)} \approx\;&\omega+
\left(
\frac{3B^2}{4m^4\omega^3}
+
\frac{8\beta}{m^2\omega^2}
\right)
\\
&+
\left(
-\frac{249}{64}\frac{B^4}{m^8\omega^7}
-\frac{45B^2\beta}{m^6\omega^6}
-\frac{144\beta^2}{m^4\omega^5}
\right).
\label{AI}
\end{split}
\tag{A1}
\end{equation}

Using Eq.~\eqref{frequen} and isolating $E$ we have

\begin{equation}
E= -\frac{m}{8}\omega^2 \mp \frac{B}{2m}. \label{A1}
\tag{A2}
\end{equation}

The equation above is substituted into Eq.~\eqref{spectro}, leading to

\begin{equation}
M= m_q+m_{\bar q}+V_0+ \frac{m}{8}\omega^2 \pm \frac{B}{2m}.
\tag{A3}
\end{equation}

Considering the parameters $\alpha$, $\beta$, $m_c$, $m_{c\bar c}$, the experimental value $M$ for the meson $c\bar c$ listed in Table~\ref{tab:ccbar_mass_spectrumI}, and substituting in Eqs.~\eqref{A1} and~\eqref{AI}, we have

\begin{equation}
\left\{
\begin{aligned}
1.888 &= \omega
+\left(
\frac{3B^2}{4m^4\omega^3}
+\frac{8\beta}{m^2\omega^2}
\right)
+\left(
-\frac{249}{64}\frac{B^4}{m^8\omega^7}
-\frac{45B^2\beta}{m^6\omega^6}
-\frac{144\beta^2}{m^4\omega^5}
\right),\\
0.456 &= 0.082525\,\omega^2
\pm 0.7573463\,B .
\end{aligned}
\right.
\tag{A4}
\end{equation}

By solving the equation system we obtain the field value $(\pm B)$ and $\omega$ for the $c\bar c$ system,

\begin{equation}
\omega \approx 1.9658,
\qquad
|B| \approx 0.1810~\mathrm{GeV}^2 .
\tag{A5}
\end{equation}

An analogous procedure can be carried out for the meson $b\bar b$.

\section{Derivation of the first-order coefficients}\label{app-B}

In this appendix, we provide the main steps used to obtain the coefficients
appearing in Eqs.~(\ref{b40})--(\ref{eq:Psi01}). We define
\begin{equation}
S=
\left[
(\hat a+\hat a^\dagger)^2
+
(\hat b+\hat b^\dagger)^2
\right],
\label{eq:S_appendix}
\end{equation}
so that the perturbation Hamiltonian can be written as
\begin{equation}
\hat H_1
=
\frac{B^2}{64m^4\omega^3}S^3
+
\frac{\beta}{m^2\omega^2}S^2.
\label{eq:H1_appendix}
\end{equation}

The state corrected up to first order is obtained from
\begin{equation}
\Psi^{(1)}_{n_1,n_2}
=
\Psi^{(0)}_{n_1,n_2}
+
\sum_{(m_1,m_2)\neq(n_1,n_2)}
\frac{
\left\langle
\Psi^{(0)}_{m_1,m_2}
\left|
\hat H_1
\right|
\Psi^{(0)}_{n_1,n_2}
\right\rangle
}{
4\alpha^{(0)}_{n_1,n_2}
-
4\alpha^{(0)}_{m_1,m_2}
}
\Psi^{(0)}_{m_1,m_2}.
\label{eq:first_order_appendix}
\end{equation}

Using the ladder-operator relations,
\begin{equation}
\hat a\,\phi_n
=
\sqrt{n}\,\phi_{n-1},
\qquad
\hat a^\dagger\phi_n
=
\sqrt{n+1}\,\phi_{n+1},
\label{eq:ladder_appendix}
\end{equation}
one obtains
\begin{equation}
(\hat a+\hat a^\dagger)^2\phi_n
=
\sqrt{n(n-1)}\,\phi_{n-2}
+
(2n+1)\phi_n
+
\sqrt{(n+1)(n+2)}\,\phi_{n+2}.
\label{eq:ladder_squared_appendix}
\end{equation}
The same relation holds for the operator $\hat b$ acting on the second
degree of freedom.

For convenience, we define
\begin{equation}
C^{(r)}_{m_1,m_2;n_1,n_2}
=
\left\langle
m_1,m_2
\left|
S^r
\right|
n_1,n_2
\right\rangle,
\qquad r=2,3.
\label{eq:C_appendix}
\end{equation}
Therefore, the matrix elements of $\hat H_1$ are
\begin{equation}
\left\langle
m_1,m_2
\left|
\hat H_1
\right|
n_1,n_2
\right\rangle
=
\frac{B^2}{64m^4\omega^3}
C^{(3)}_{m_1,m_2;n_1,n_2}
+
\frac{\beta}{m^2\omega^2}
C^{(2)}_{m_1,m_2;n_1,n_2}.
\label{eq:matrix_H1_appendix}
\end{equation}

From Eq.~(\ref{eq:first_order_appendix}) and
\[
4\alpha^{(0)}_{n_1,n_2}
=
(n_1+n_2+1)\omega,
\]
the energy-denominator factor can be written as
\begin{equation}
4\alpha^{(0)}_{n_1,n_2}
-
4\alpha^{(0)}_{m_1,m_2}
=
-\Delta N\,\omega,
\qquad
\Delta N
=
m_1+m_2-n_1-n_2.
\label{eq:DeltaN_appendix}
\end{equation}
For the transitions retained in the main text,
$m_1+m_2>n_1+n_2$, and therefore $\Delta N>0$.

Consequently, each first-order coefficient can be written as
\begin{equation}
A_{m_1,m_2;n_1,n_2}
=
-\frac{1}{\omega}
\left[
\frac{B^2}{64m^4\omega^3}
\frac{
C^{(3)}_{m_1,m_2;n_1,n_2}
}{
\Delta N
}
+
\frac{\beta}{m^2\omega^2}
\frac{
C^{(2)}_{m_1,m_2;n_1,n_2}
}{
\Delta N
}
\right].
\label{eq:A_appendix}
\end{equation}

For the ground state $(n_1,n_2)=(0,0)$, the contributions retained
in the main-text expression are
\begin{equation}
\begin{array}{c|c|c|c}
(m_1,m_2) & \Delta N & C^{(2)} & C^{(3)} \\
\hline
(2,0) & 2 & 8\sqrt{2}  & 72\sqrt{2} \\
(2,2) & 4 & 4           & 72 \\
(4,0) & 4 & 2\sqrt{6}   & 36\sqrt{6} \\
(4,2) & 6 & 0           & 12\sqrt{3} \\
(6,0) & 6 & 0           & 12\sqrt{5}
\end{array}
\label{eq:ground_coefficients_appendix}
\end{equation}
Substituting these values into Eq.~(\ref{eq:A_appendix}) yields the
coefficients appearing in Eq.~(\ref{b40}).

For the first excited state $(n_1,n_2)=(1,0)$, the corresponding
coefficients retained in the main-text expression are
\begin{equation}
\begin{array}{c|c|c|c}
(m_1,m_2) & \Delta N & C^{(2)} & C^{(3)} \\
\hline
(1,2) & 2 & 12\sqrt{2}  & 144\sqrt{2} \\
(1,4) & 4 & 2\sqrt{6}   & 48\sqrt{6} \\
(3,0) & 2 & 12\sqrt{6}  & 144\sqrt{6} \\
(3,2) & 4 & 4\sqrt{3}   & 96\sqrt{3} \\
(3,4) & 6 & 0           & 36 \\
(1,6) & 6 & 0           & 12\sqrt{5} \\
(5,0) & 4 & 2\sqrt{30}  & 48\sqrt{30} \\
(5,2) & 6 & 0           & 12\sqrt{15} \\
(7,0) & 6 & 0           & 12\sqrt{35}
\end{array}
\label{eq:excited_coefficients_appendix}
\end{equation}

For the state $(n_1,n_2)=(0,1)$, the coefficients follow analogously
by the exchange $n_1\leftrightarrow n_2$, leading to
Eq.~(\ref{eq:Psi01}).

\section{Sensitivity analysis for the magnetic field}\label{app-C}

In this appendix we estimate the sensitivity of the magnetic field $B$ for the $c\bar c$ system. The values of $B$ and $\omega$ are not chosen arbitrarily. These values are obtained by solving the coupled system defined by the energy-spectrum equation and the mass-spectrum equation.

For the $c\bar c$ ground state, we define

\begin{align}
F_1(\omega,B)
=&
\omega
+
\left(
\frac{3B^2}{4m^4\omega^3}
+
\frac{8\beta}{m^2\omega^2}
\right)
+
\left(
-\frac{249}{64}\frac{B^4}{m^8\omega^7}
-\frac{45B^2\beta}{m^6\omega^6}
-
\frac{144\beta^2}{m^4\omega^5}
\right)
-
4\alpha,
\end{align}
and
\begin{equation}
F_2(\omega,B)
=
\frac{m}{8}\omega^2
+
\frac{B}{2m}
-
\left(
M-m_c-m_{\bar c}-V_0
\right).
\end{equation}

The values are obtained from
$
F_1(\omega,B)=0,
\qquad
F_2(\omega,B)=0.
$
For the central values used in the text,
$
\alpha=0.472,
\,\,
\beta=0.191~{\rm GeV}^2,
\,\,
m_c=m_{\bar c}=1.3205~{\rm GeV},
\,\,
m=0.6602~{\rm GeV},
\,\,
V_0=0,
\,\,
M=3.097~{\rm GeV},
$
we obtain
\begin{equation*}
\omega_0=1.9658,
\qquad
B_0=0.1810~{\rm GeV}^2.
\end{equation*}

In order to estimate how the magnetic field changes when an input parameter $p_i$ is varied, we linearize the system around the central solution:
\begin{equation}
\begin{pmatrix}
\delta F_1 \\
\delta F_2
\end{pmatrix}
=
\begin{pmatrix}
F_{1,\omega} & F_{1,B} \\
F_{2,\omega} & F_{2,B}
\end{pmatrix}
\begin{pmatrix}
\delta \omega \\
\delta B
\end{pmatrix}
+
\begin{pmatrix}
F_{1,p_i} \\
F_{2,p_i}
\end{pmatrix}
\delta p_i .
\end{equation}

Since the solution must continue satisfying $F_1=F_2=0$, one has $\delta F_1=\delta F_2=0$. Therefore,
\begin{equation*}
\begin{pmatrix}
\delta \omega \\
\delta B
\end{pmatrix}
=
-
J^{-1}
\begin{pmatrix}
F_{1,p_i} \\
F_{2,p_i}
\end{pmatrix}
\delta p_i,
\end{equation*}
where
\begin{equation*}
J=
\begin{pmatrix}
F_{1,\omega} & F_{1,B} \\
F_{2,\omega} & F_{2,B}
\end{pmatrix}.
\end{equation*}

Then the local sensitivity of the magnetic field is
\begin{equation}
\frac{\partial B}{\partial p_i}
=
\frac{
F_{2,\omega}F_{1,p_i}
-
F_{1,\omega}F_{2,p_i}
}{
F_{1,\omega}F_{2,B}
-
F_{1,B}F_{2,\omega}
}.
\end{equation}

Considering the central solution, the Jacobian matrix is
\begin{equation}
J=
\begin{pmatrix}
2.6307 & -0.4857 \\
0.3244 & 0.7573
\end{pmatrix},
\end{equation}
with determinant
$
\det J = 2.1499.
$

The resulting derivatives are given by
$$
\begin{matrix}
\frac{\partial B}{\partial M} &= +1.2236,
\,\,
\frac{\partial B}{\partial m_c}
=
\frac{\partial B}{\partial m_{\bar c}}
&= -1.2236,
\,\,
\frac{\partial B}{\partial V_0} &= -1.2236,
\,\,
\frac{\partial B}{\partial m} &= +0.1768,
\,\,
\frac{\partial B}{\partial \alpha} &= -0.6037,
\,\,
\frac{\partial B}{\partial \beta} &= -0.8184.
\end{matrix}
$$

Since in the $c\bar c$ system $m_c=m_{\bar c}$, a simultaneous variation of
the quark and antiquark masses gives
\begin{equation}
\left.
\frac{\partial B}{\partial m_c}
\right|_{m_c=m_{\bar c}}
=
-2.4472.
\end{equation}

The variation in $B$ is then estimated from
\begin{equation}
\Delta B_i
\simeq
\left|
\frac{\partial B}{\partial p_i}
\right|
\Delta p_i .
\end{equation}

If all input variations are assumed to be independent, the total sensitivity
estimate is
\begin{equation}
\sigma_B^{\rm sens}
=
\sqrt{
\sum_i
\left(
\frac{\partial B}{\partial p_i}\Delta p_i
\right)^2
}.
\end{equation}

Using the variations shown in Table~\ref{tab:sensitivity_ccbar}, we obtain the local sensitivity of the magnetic field for the $c\bar c$ system.

\end{document}